\documentclass[conference]{IEEEtran}
\usepackage{graphicx}
\usepackage{amssymb,latexsym}
\usepackage{amsmath}
\usepackage{lipsum}
\usepackage{empheq}
\usepackage{makecell}
\usepackage{epsfig}
\usepackage{multirow}
\usepackage{subfig}
\usepackage{epstopdf}  
\usepackage{color}

\usepackage{array}
\usepackage{enumitem}
\setlist[itemize]{align=parleft,left=0pt..1em}

\usepackage[all]{background}

\usepackage{stackengine}
\setstackEOL{\\}
\setstackgap{L}{\normalbaselineskip}
\SetBgContents{\color{gray}{\tiny \Longstack{PREPRINT - accepted by 30th IFIP/IEEE International Conference on Very Large Scale Integration 2022 (VLSI-SoC 2022)}}}
\SetBgPosition{3.5cm,1cm}
\SetBgOpacity{1.0}
\SetBgAngle{0}
\SetBgScale{1.8}

\usepackage[]{float,latexsym,amssymb,amsfonts,amstext,times,multirow}

\def\BibTeX{{\rm B\kern-.05em{\sc i\kern-.025em b}\kern-.08em
    T\kern-.1667em\lower.7ex\hbox{E}\kern-.125emX}}

\title{PA-PUF: A Novel Priority Arbiter PUF\vspace{-5mm}}  
\date{}
\author{\IEEEauthorblockN{Simranjeet Singh\IEEEauthorrefmark{1}, Srinivasu Bodapati\IEEEauthorrefmark{2}, Sachin Patkar\IEEEauthorrefmark{1}, 
Rainer Leupers\IEEEauthorrefmark{4}, \\
Anupam Chattopadhyay\IEEEauthorrefmark{3}, Farhad Merchant\IEEEauthorrefmark{4} \IEEEauthorblockA{\IEEEauthorrefmark{1}Indian Institute of Technology, Bombay, \IEEEauthorrefmark{2}Indian Institute of Technology, Mandi,\\ \IEEEauthorrefmark{3}Nanyang Technological University, Singapore, \IEEEauthorrefmark{4}RWTH Aachen University, Germany}}\{simranjeet, patkar\}@ee.iitb.ac.in, srinivasu@iitmandi.ac.in, anupam@ntu.edu.sg, \{leupers, merchantf\}@ice.rwth-aachen.de \vspace{-7mm}}

\begin{document}
\bstctlcite{IEEEexample:BSTcontrol} 
\maketitle

\begin{abstract}  %
This paper proposes a 3-input arbiter-based novel physically unclonable function (PUF) design. Firstly, a 3-input priority arbiter is designed using a simple arbiter, two multiplexers (2:1), and an XOR logic gate. The priority arbiter has an equal probability of 0's and 1's at the output, which results in excellent uniformity (49.45$\%$) while retrieving the PUF response. Secondly, a new PUF design based on priority arbiter PUF (PA-PUF) is presented. The PA-PUF design is evaluated for uniqueness, non-linearity, and uniformity against the standard tests. The proposed PA-PUF design is configurable in challenge-response pairs through an arbitrary number of feed-forward priority arbiters introduced to the design. We demonstrate, through extensive experiments, reliability of 100$\%$ after performing the error correction techniques and uniqueness of 49.63$\%$. 
Finally, the design is compared with the literature to evaluate its implementation efficiency, where it is clearly found to be superior compared to the state-of-the-art.       
\end{abstract}



\section{Introduction}
\label{sec:Intro}

Physical unclonable functions (PUFs) have great prominence in today's secure device authentication and secure communication~\cite{LeeDevadas2004}. A PUF takes advantage of randomness due to manufacturing process variation in integrated circuits and generates a unique response for every device by tapping into the sources of entropy such as variations in path delay or timing~\cite{LeeDevadas2004}. An integrated circuit within the PUF maps the input challenge with a unique response, thus creating a challenge-response pair (CRP). These CRPs are utilized to design various security protocols, ranging from device attestation to data encryption. The concept of a PUF was first reported in~\cite{Pappu2002}. Based on the parameters of how the response is derived, various kinds of PUFs have been proposed in the literature. These parameters can be delay, memory, or the difference between the current in the rails. In the literature, PUF designs are categorized as delay-based and memory-based PUF - of which the prominent examples are Arbiter PUF~\cite{LimDevadas2005a}, ring oscillator PUF~\cite{SuhDevadas2007a}, SRAM PUF~\cite{Guajardo2007a}, Butterfly PUF~\cite{kumar2008a}, Glitch PUF~\cite{Suzuki2010a} and MEmory Cell-based Chip Authentication (MECCA) PUF~\cite{Krishna2011a}. These advanced PUFs are being threatened with various attacks~\cite{pufml2, pufml3}. Therefore, we need to keep studying new constructions of PUFs that potentially thwart such attacks with minimalistic~overhead.

Arbiter PUF (APUF) is one of the delay-based PUF designs, where an arbiter decides the response of a PUF between the two data paths comprised of cross-coupled multiplexers~\cite{Ganta2013}. For this purpose, the arbiter PUF uses the analog timing difference between the two data paths and decides the output based on the timing difference between the two lines. Several modifications to the classical arbiter PUF are presented in the literature. In arbiter PUF, it is possible to predict the relation between the challenge and response through software modeling and programs. To prevent the modeling-based attacks, various modifications were proposed, such as multi-arbiter PUF~\cite{Klybik2015},~\cite{zalivaka2016a} and double arbiter PUF~\cite{Machida2014a}. The multi-arbiter PUF design consists of arbiters in each multiplexer stage and a multiplexer to choose the arbiter response. An alternative is to take the PUF response as XOR of the arbiter outputs, which improves the uniqueness, reliability, and robustness of the PUF. 
 
The arbiter is an electronic circuit that can identify a signal's first occurrence. A simple D flip-flop can be used as an arbiter where one signal is connected as the clock and the other as the data signal. Priority arbiter is typically used in the Network-On-Chip (NoC)~\cite{Dimitra2012}, in order to determine the priority of the data request among the several requests~\cite{Dimitra2012}. Different kinds of arbiters are proposed in literature~\cite{Wang2009}, such as daisy chain arbiter, round-robin arbiter, and dynamic arbiter. Based on the application of the NoC, the arbiters are designed. The concept of priority arbiter from the NoC communication can be adopted into the PUF to improve the design efficiency. The advantage of having a priority arbiter is that the design has more non-linearity compared to the simple arbiter PUF. In this paper, a new priority arbiter is proposed, which demonstrates good uniformity. Based on that, a novel PA-PUF is designed. The major contributions of this work are reported as follows:


\begin{itemize}
\item A new PUF using the priority arbiter called PA-PUF is designed. The PA-PUF offers a uniqueness of 49.63 $\%$ and uniformity of 49.45$\%$ at the output. The non-linearity in the output of the PUF is increased with the use of a priority arbiter.

\item We demonstrate the configurability of PA-PUF by varying the number of CRPs. The number of CRPs can be increased by increasing the length of the data path by introducing more feed-forward arbiters.


\item The performance of the proposed priority arbiter PUF is studied as a function of the number of feed-forward arbiters in the data path and the length of the data path. For example, the uniqueness of the PUF can be increased by increasing the length of the data path. It offers a reliability of 94.5$\%$ for a 128-bit response, which can be increased to 100$\%$ by implementing Bose-Chaudhuri-Hocquenghem (BCH) error-correcting codes~\cite{BOSE196068}.
\end{itemize}

\begin{figure}[!t]
\centering
\includegraphics[width=0.42\textwidth]{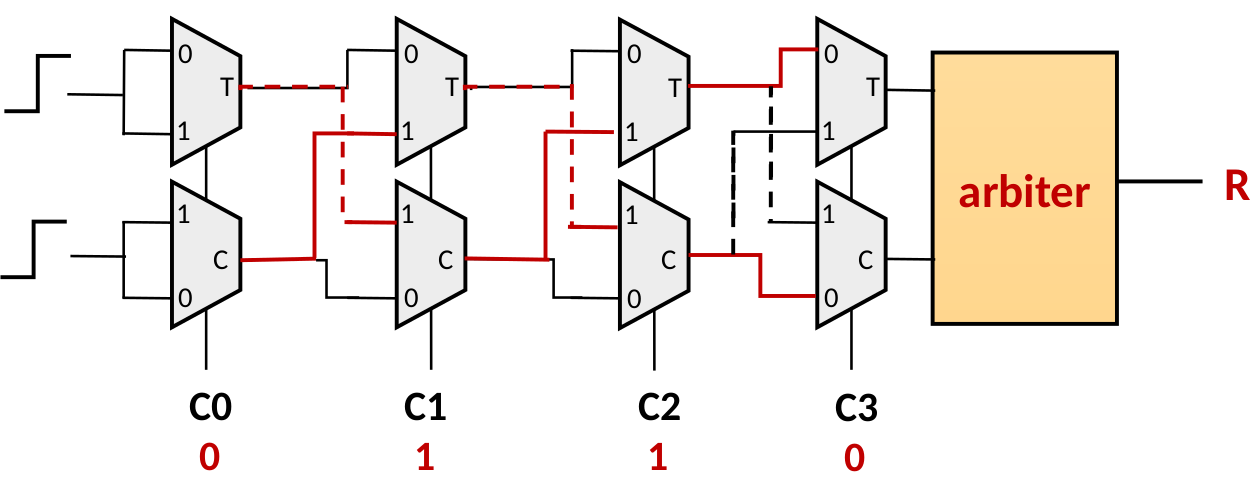}
\caption{Classical arbiter PUF design, red lines indicate the data path of a given challenge.}
\label{fig:arbiter_PUF}
\end{figure}


The rest of the paper is structured as follows. Section~\ref{sec:PAPUF} provides the design  
 insights of the newly proposed PUF. The experimental results are given in section~\ref{sec:Expt_results}.
  Section~\ref{sec:comp} compares the proposed design with state-of-the-art designs from literature. Finally,  Section~\ref{sec:conclusions} provides conclusions and outlook of this work.


\section{Proposed Priority Arbiter PUF}
\label{sec:PAPUF}
  
  \begin{figure}[!t]
\centering
\includegraphics[scale=0.48]{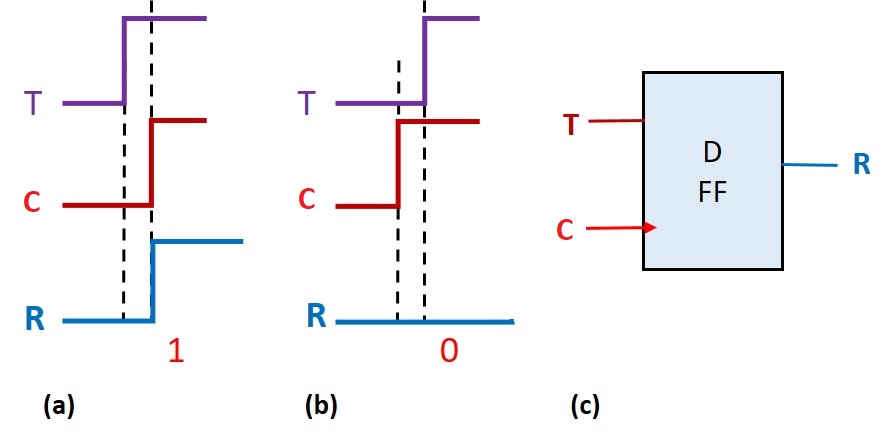}
\caption{(a), (b) possible input wave forms, and (c) D flip-flop.}
\label{fig:arbiter_ckt}
\end{figure}

A classical arbiter PUF with multiplexers and an arbiter is depicted in Fig.~\ref{fig:arbiter_PUF}. The arbiter PUF is a delay-based design and the delay difference is taken from the cross-coupled multiplexers as shown in the design. 
The challenge of the arbiter PUF is given as select signals of the multiplexers in the data path, a low to high 
transition given at the input of the multiplexer will be propagated to the arbiter in the path selected by
the challenge of the PUF. 
The arbiter is implemented using a D flip-flop, and the same is presented with the possible operations in Fig.~\ref{fig:arbiter_ckt}, with two signals as Top ($T$) and Center ($C$). When the signal $C$ arrives first, then the output of the arbiter is led to `0'; otherwise, the output is `1'. The same is explained in the Fig.~\ref{fig:arbiter_ckt} (a) $\&$ (b). Fig.~\ref{fig:arbiter_PUF} presents the symmetric path of the circuit through the multiplexers.
The multiplexers' select signals (challenge) generate different delay paths, resulting in a unique response for every combination. The major disadvantage with the arbiter PUF is that the uniqueness of the PUF is less. There are several modifications to the classical arbiter PUF by adding a feed-forward path~\cite{Parhi2014} in the design to introduce the non-linearity in the results.

In arbiter PUF, it is possible to predict the relation between the challenge and response through software modeling and programs. To prevent this modeling prediction, various modifications have been proposed in the past. One alternative of arbiter PUF is multi-arbiter PUF~\cite{Klybik2015},~\cite{zalivaka2016a} and double arbiter PUF~\cite{Machida2014a}. All these methods are
proposed to improve the uniqueness, reliability, and robustness of the PUF output. Since the order of non-linearity in the arbiter is less, so by considering multi-arbiter designs, the non-linearity can be increased in the design. Next, we will propose a priority arbiter-based PUF design.


 \begin{figure}[!t]
   \subfloat[\label{1a}]{%
       \includegraphics[width=0.27\linewidth]{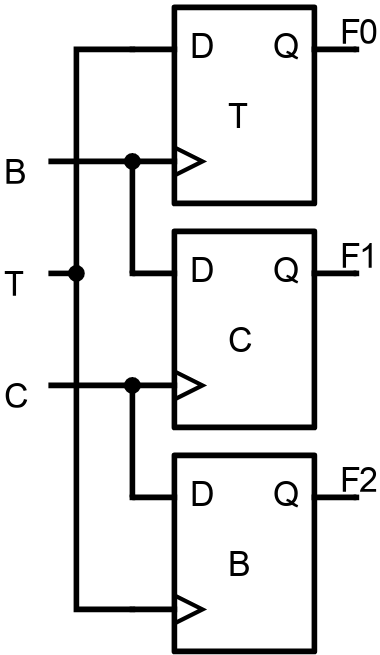}}
    \hfill
  \subfloat[\label{1b}]{%
        \includegraphics[width=0.55\linewidth]{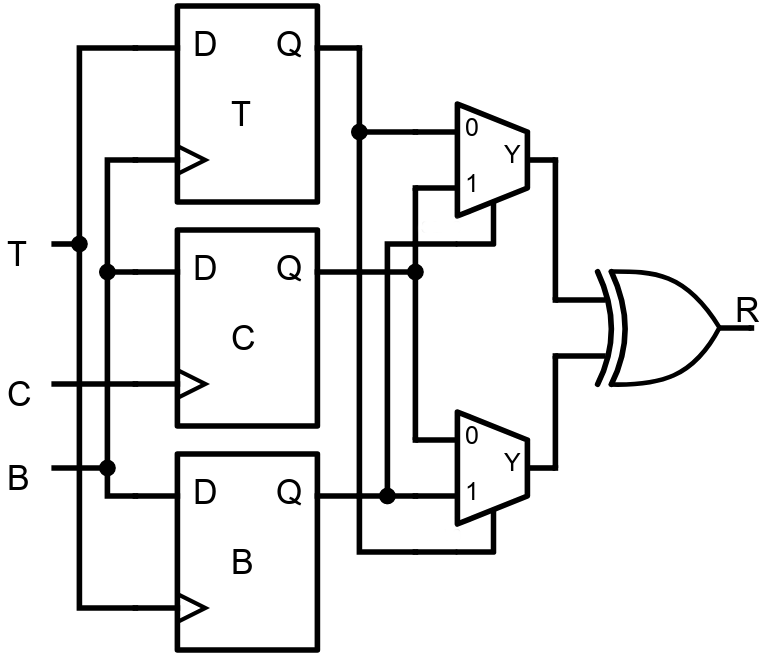}}
    \\
  \caption{ (a) Feed-forward arbiter for the proposed PA-PUF; T, C and B as the top, center and bottom data lines; $F0, F1$ and $F2$ are the three outputs from the feed-forward arbiter (b) Three-input priority arbiter; T, C and B as the top, center and bottom data lines and R as response bit.\vspace{-3mm}}
\label{fig:prio_arbiter}
  
 \end{figure}


A new three-input priority arbiter is proposed to decide the response as per the priority of the input's arrival time. The arbiter used in the classical arbiter is a two-input circuit. Now, an extra input is added to increase the non-linearity in the output. The proposed three-input priority arbiter diagram is shown in Fig.~\ref{1b}, which is designed with three D-flip flops, two 2-to-1 multiplexers, and an XOR logic gate. The three inputs, T, C, and B ({\em top, center, and bottom}) are given to the D flip-flops. Next, the select signal and the data line of the multiplexer are taken from the outputs of the D flip-flops. Finally, the output of multiplexers is applied to the input of the XOR gate to generate a response bit.

\begin{figure}[!t]
\centering
\includegraphics[width=0.48\textwidth]{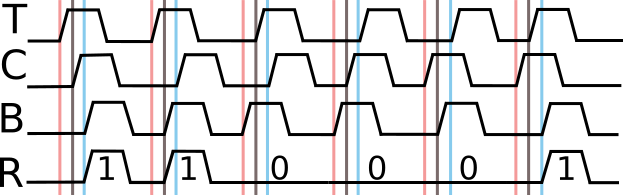}
\caption{Priority arbiter possible output (R) combinations with T, C and B as the top, center and bottom data lines. \vspace{-3mm}}


\label{fig:prio_arbiter_wave}
\end{figure}

The operation of the proposed priority arbiter is shown in Fig.~\ref{fig:prio_arbiter_wave}, where the output is decided based on the priority of the input arrival times. Some of the possible conditions are considered in the plot. Consider a case where the arrival times of these signals are as follows, T, C, and B (first case in Fig.~\ref{fig:prio_arbiter_wave}). Since T comes first in time, the outputs of the bottom and center D flip-flops (DFF B and DFF C) are `0' and `0', while the top D flip-flop (DFF T) output is `1'. After multiplexer output, the input of the XOR gate will be `1' and `0'. This results in the output of the arbiter being `1'. The Fig.~\ref{fig:prio_arbiter_wave} demonstrates the possible conditions
$\{T,B,C\},\{T,C,B\},\{B,T,C\},\{B,C,T\},\{C,B,T\}$ and $\{C,T,B\}$  out of these six conditions, the output is `0' in three cases and `1' in  three cases. This further results in achieving good uniformity in the output of the priority arbiter. The same circuit with different connections of the inputs will~lead to non-uniform outputs (with non-equal probability of 0's and~1's).

\begin{figure}[!t]
\includegraphics[width=0.48\textwidth]{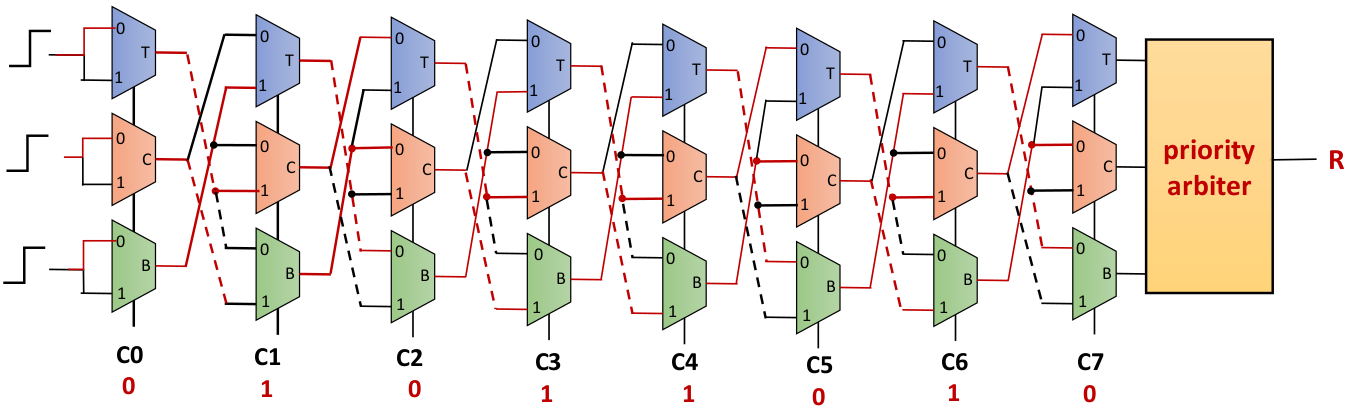}
\caption{Schematic of proposed PA-PUF which includes three parallel multiplexer lines (T, C, and B) and a priority arbiter at the end to generate the response bit.\vspace{-3mm}}
\label{fig:prio_arbiter_PUF}
\end{figure}

\begin{figure}[!t]
 \includegraphics[width=0.48\textwidth]{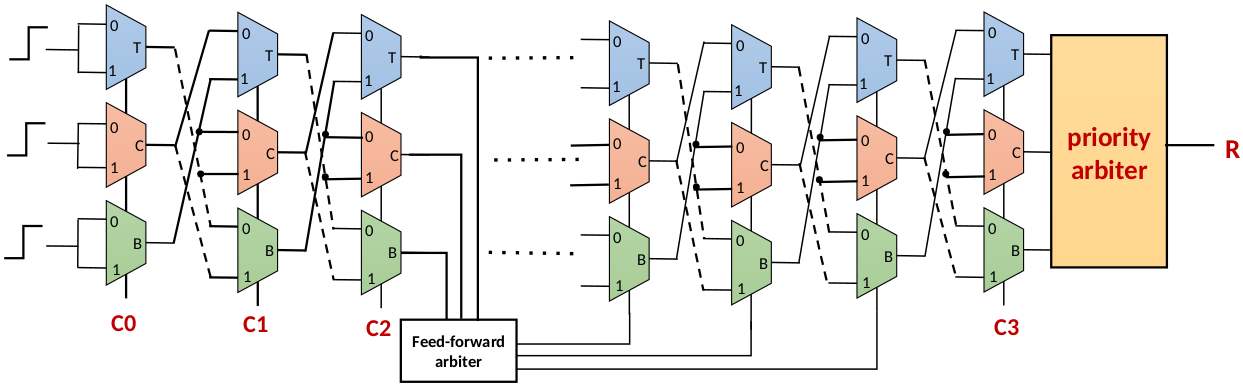} 
\caption{ Proposed PA-PUF with feed-forward arbiter.\vspace{-3mm}}
\label{fig:PAPUF_FF}
\end{figure}

The PA-PUF is presented in Fig.~\ref{fig:prio_arbiter_PUF}, where the modifications to the classical arbiter PUF have been done by adding a third data path in addition to the two data paths. Fig.~\ref{fig:prio_arbiter_PUF} shows the working of the PA-PUF for a given challenge (marked in red). This shows that just by increasing the hardware by one-third, the non-linearity in the output can be increased to an extreme. In the case of arbiter PUF, when the output is `1', then it can be understood that signal $T$  arrives before signal $C$. While in the case of priority arbiter PUF, as explained in Fig.~\ref{fig:prio_arbiter_wave}, the output is `1', in three cases. Hence, it is difficult to find which signal comes first. Moreover, the design is extended by introducing the feed-forward paths to increase the robustness in the design. The priority arbiter PUF with the feed-forward path is shown in Fig.~\ref{fig:PAPUF_FF}, while the feed-forward arbiter for the priority arbiter PUF is shown in Fig.~\ref{1b}.    


 \section{Experimental results}
\label{sec:Expt_results}

The proposed PA-PUF is designed using Verilog programming and implemented on Nexys Video Artix-7 FPGA board. 
The responses of the PUF are collected using the universal asynchronous receiver/transmitter (UART) protocol. 
Next, we will discuss the proposed PA-PUF (with a feed-forward arbiter) and the implications of deriving a security key. For a good PUF design, it should satisfy some important criteria such as {\em intra-chip Hamming distance}, {\em inter-chip Hamming distance}. Based on the results of the inter-chip and intra-chip Hamming distance, we can analyse the other important parameters such as {\em uniformity}, {\em robustness}, {\em uniqueness}, {\em bit-aliasing}, and {\em reliability}. Intra Hamming distance (HD) is the Hamming distance between the responses of the PUF design within the chip. Ideally, a single bit change in the challenge should result in a 50\% Hamming distance in the responses bits. The intra HD can be calculated by the formula given in~equation~\ref{eq:HD}.



 \begin{equation}
Intera\ HD =  \sum_{i=1}^k\frac{HD(R_i,R_{i+1})}{n} \times 100
 \label{eq:HD}
 \end{equation}
 
\begin{figure}[!t]
\centering
\includegraphics[scale=0.4]{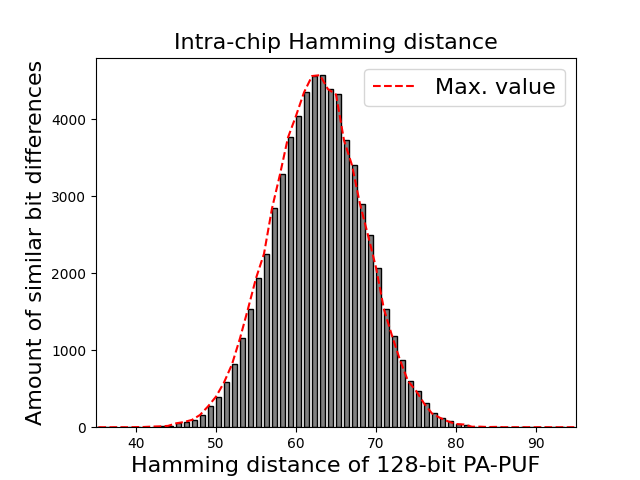}
\caption{Intra-chip Hamming distance plot of 128-bit PA-PUF. \vspace{-3mm}}
\label{fig:PAPUF_HD16}
\end{figure}

Here, `k' is the total number of challenges given to the PUF, $R_i$ and $R_{i+1}$ are
the responses to the challenges $C_i$ and $C_{i+1}$ respectively. The challenges are differed by one-bit change, and Fig.~\ref{fig:PAPUF_HD16} shows the Hamming distance in responses of the 128-bit PA-PUF. Fig.~\ref{fig:PAPUF_HD16} also shows the HD has a maximum at the half of the responses, and the plot follows the Gaussian distribution.
 


\begin{figure}[!t]
\centering
\includegraphics[scale=0.4]{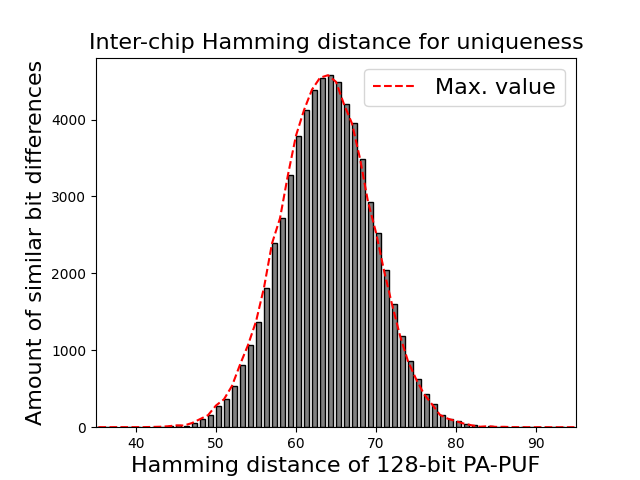}
\caption{Inter-chip Hamming distance plot of 128-bit PA-PUF.\vspace{-3mm}}
\label{fig:PAPUF_IHD16}
\end{figure}

Inter HD is the Hamming distance of the responses between two chips of the same family. We have used three FPGAs of the same family/configuration to produce the results on inter HD. Fig.~\ref{fig:PAPUF_IHD16} shows the inter HD between responses of two FPGA boards, which is calculated by the formula given in equation~\ref{eq:IHD}.
  
  \begin{equation}
Inter\ HD = \frac{2}{k(k-1)}\sum_{i=1}^{k-1}\sum_{j=i+1}^k\frac{HD(R_i,R_j)}{n} 
 \label{eq:IHD} 
 \end{equation}

Where, $i$ and $j$ are two different FPGAs. $R_i$ and $R_j$ are the responses from $FPGA_i$ and $FPGA_j$ for the challenge C respectively. $k$ is number of PUF designs (3 in our case).


\begin{figure}[!t]
\centering
\includegraphics[scale=0.45]{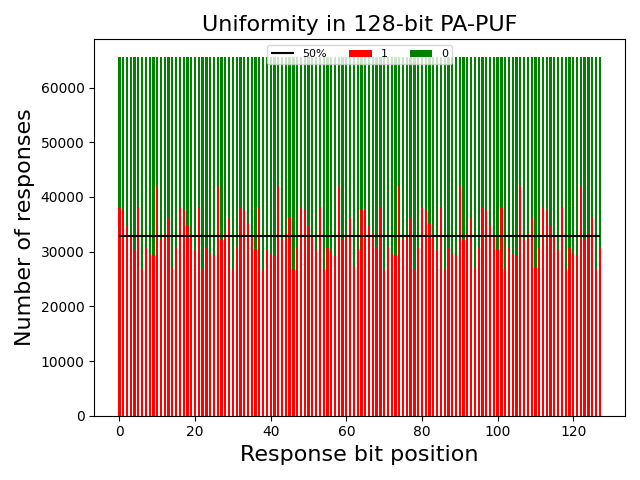}
\caption{Uniformity of 128-bit PA-PUF along with 50\% probability line. \vspace{-3mm}}
\label{fig:PAPUF_uni}
\end{figure}



\subsection{Robustness}
Any response bit of the PUF should not be stuck at logic `0' or `1'.
The stable `0', stable `1' and unstable bits are calculated for various sizes of the PA-PUF design. The results are given in Table~\ref{tab:robustness} and the results reveal that the design has good robustness.

    \begin{table}[!b]
    \scriptsize
   \centering
         \caption{Robustness results of stable `0' , `1' and unstable~bits.} 
   \begin{tabular}{|c|c|c|c|c|} \hline
   \textbf{Board}  & \textbf{Number of}        & \textbf{Stable `0'}        & \textbf{Stable `1'}       &    \textbf{Unstable}   \\    
    \textbf{Artix-7} &  \textbf{bits}                         &  \textbf{\%}         & \textbf{\%}       & \textbf{\%}    \\         \hline 
    \multicolumn{5}{|c|}{\textbf{8-bit proposed PA-PUF}} \\ \hline
     1  & 0.5  x $10^6$ & 25.93   & 24.64  & 49.43  \\ \hline
     2  & 0.5  x $10^6$ &  27.65  & 21.71  & 50.64  \\ \hline
    \multicolumn{5}{|c|}{\textbf{16-bit proposed PA-PUF}} \\ \hline
     1  & 1  x $10^6$ & 26.65   & 24.24  & 49.11  \\ \hline
     2  & 1  x $10^6$ &  25.07  & 25.00  & 49.93  \\ \hline
     \multicolumn{5}{|c|}{\textbf{32-bit proposed PA-PUF}} \\ \hline
     1  & 2  x $10^6$ & 26.24   & 24.92  & 48.84  \\ \hline
     2  & 2  x $10^6$ &  27.66  & 23.19  & 49.15  \\ \hline
     \multicolumn{5}{|c|}{\textbf{64-bit proposed PA-PUF}} \\ \hline
     1  & 4  x $10^6$ & 30.52   & 21.32  & 48.16  \\ \hline
     2  & 4  x $10^6$ &  28.03  & 23.01  & 48.96  \\ \hline
     \multicolumn{5}{|c|}{\textbf{128-bit proposed PA-PUF}} \\ \hline
     1  & 8  x $10^6$ & 31.80   & 19.79  & 48.41 \\ \hline
     2  & 8  x $10^6$ &  30.49  & 20.58  & 48.93  \\ \hline   
     \multicolumn{5}{|c|}{\textbf{128-bit Ring oscillator PUF}} \\ \hline
   1          &   8 $\times 10^6$     & 28.1       &  23.3       &    48.6 \\    \hline
   2         &   8 $\times 10^6$      & 27.6       &  23.6      &     48.8 \\    \hline
   \textbf{Mean} & ~                               &  \textbf{27.85}     &  \textbf{23.45} &   \textbf{48.7}  \\  \hline
   \textbf{Mean}~\cite{GuNeill2017a}  & ~                  &  \textbf{29.51}             &  \textbf{30.25}         &  \textbf{40.22}           \\ \hline 
           \end{tabular}
       \label{tab:robustness}
   \end{table}


\subsection{Uniformity and Bit-aliasing}
  
The response should have an equal number of  0`s and 1`s. This can be calculated
using the `uniformity' of the PUF. The ideal value of uniformity is 50$\%$, which means  the PUF response is uniformly distributed. Fig.~\ref{fig:PAPUF_uni} shows the distribution of 0`s and 1`s in response bits. Table~\ref{tab:uniform_bitalias} gives the values for various sizes of  the PUF response. 
The other parameter of interest is the `bit-aliasing', which is calculated over different chips/devices. The experiment is conducted on three Nexys Video Artix-7 FPGA boards of the same family. The PUF design is placed in the exact physical locations across different boards to find whether any particular bit position is permanently connected to the logic `0' or `1' irrespective of the challenge or a board. The ideal value of the bit aliasing is 50$\%$, which is near to the values recorded in Table~\ref{tab:uniform_bitalias}. 

    \begin{table}[!b]
    \scriptsize
\centering
    \caption {Uniformity and bit-aliasing for the proposed PUF.}
    \begin{tabular}{|c|c|c|}
    \hline
    \textbf{Parameter} & \textbf{Uniformity}    &\textbf{Bit-alising}          \\ \hline
    \textbf{Ideal value} &  \textbf{ 50 }\%     & \textbf{50 \%}  \\ \hline
    \multicolumn{3}{|c|}{\textbf{8-bit proposed PA-PUF}}  \\ \hline
    Maximum   &100$\%$     & 52.41 $\%$           \\ \hline
    Minimum   & 0 $\%$        & 42.74 $\%$            \\ \hline
    Average   & 49.3 $\%$         & 49.17 $\%$        \\ \hline 
     \multicolumn{3}{|c|}{\textbf{16-bit proposed PA-PUF}} \\ \hline
    Maximum   & 100 $\%$     & 64.37 $\%$        \\ \hline
    Minimum   & 0 $\%$        & 39.35 $\%$            \\ \hline
    Average   & 49.25 $\%$         & 48.77 $\%$        \\ \hline
     \multicolumn{3}{|c|}{\textbf{32-bit proposed PA-PUF}} \\ \hline
    Maximum   & 84.37 $\%$     &60.79 $\%$        \\ \hline
    Minimum   & 9 $\%$        & 34.87 $\%$            \\ \hline
    Average   & 47.8 $\%$         & 47.76 $\%$        \\ \hline
       \multicolumn{3}{|c|}{\textbf{64-bit proposed PA-PUF}} \\ \hline
    Maximum   & 78.75 $\%$     & 54.82 $\%$        \\ \hline
    Minimum   & 21.87 $\%$        & 32.39 $\%$            \\ \hline
    Average   & 47.47 $\%$         & 47.41 $\%$        \\ \hline
        \multicolumn{3}{|c|}{\textbf{128-bit proposed PA-PUF}} \\ \hline
    Maximum   & 67.9 $\%$     & 51.69 $\%$        \\ \hline
    Minimum   & 25 $\%$        & 33.03 $\%$            \\ \hline
    Average   & 49.45 $\%$         & 49.66 $\%$        \\ \hline
    \multicolumn{3}{|c|}{\textbf{128-bit Ring oscillator PUF}} \\ \hline
    Maximum   & 67.2 $\%$     & 52.82 $\%$        \\ \hline
    Minimum   & 0 $\%$        & 0 $\%$            \\ \hline
    Average   & 48.3 $\%$         & 47.8 $\%$        \\ \hline
       \end{tabular}
   \label{tab:uniform_bitalias}
\end{table}

\subsection{Reliability and Uniqueness}

  \begin{figure}[!t]
\centering
\includegraphics[scale=0.4]{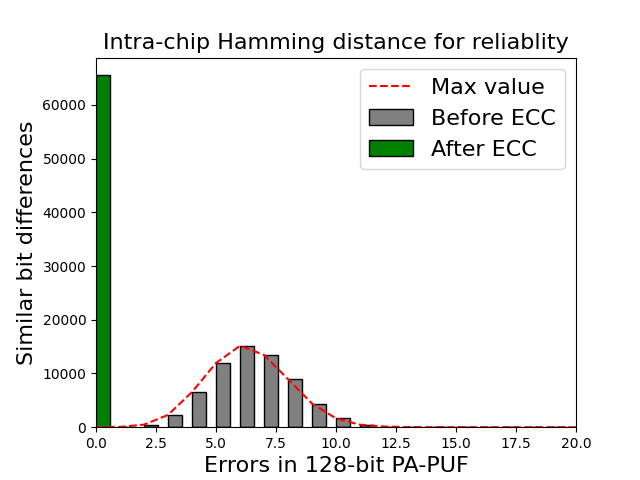} \\ 
\caption{Reliability plot of 128-bit priority arbiter PUF.}
\label{fig:PAPUF_Rel}
\end{figure}

Reliability and uniqueness are the most important metrics in deciding the design of the PUF. The 
uniqueness is defined as how uniquely the design can identify from one chip to the other chip of the same family. The ideal value of the uniqueness is 50$\%$; hence the response of the PUF from one chip to the other is differed by half of its length. Uniqueness is calculated using the `inter Hamming distance' of the response, and the calculated values are tabulated in Table~\ref{tab:uniq_rel_PAPUF} for various sizes of the proposed PUF. Further, the reliability is also given in Table~\ref{tab:uniq_rel_PAPUF}, which is calculated using the `intra-Hamming distance' of the response by giving the same challenge over million times, and Fig.~\ref{fig:PAPUF_Rel} depicts the Hamming distance plot to calculate the reliability. The error that occurred in the response of the PUF overages can be corrected using the error correction mechanisms~\cite{Dodis2014a} and~\cite{maes2012a}. It has been shown in Fig.~\ref{fig:PAPUF_Rel} that after BCH error correction codes, reliability can increase from 94.5\% to 100\% in PA-PUF.

\begin{table}[!b]
\scriptsize
   \centering
    \caption{Uniqueness and Reliability for various sizes of the proposed PA-PUF.}
    \begin{tabular}{|c|c|c|}
    \hline
    \textbf{PUF size} & \textbf{Uniqueness} & \textbf{Reliability} \\ \hline
    8-bit    & 50   $\%$      & 99.45 $\%$          \\ \hline
    16-bit   & 50 $\%$         & 99.05 $\%$       \\ \hline
    32-bit   & 50  $\%$       & 98.84 $\%$           \\ \hline
    64-bit   & 48.4 $\%$       & 98.15 $\%$       \\ \hline
    128-bit  & 49.63 $\%$      & 95.37$\%$          \\ \hline
    \end{tabular}
   
    \label{tab:uniq_rel_PAPUF}
\end{table}

   \begin{figure}[!t]
\centering
\includegraphics[scale=0.5]{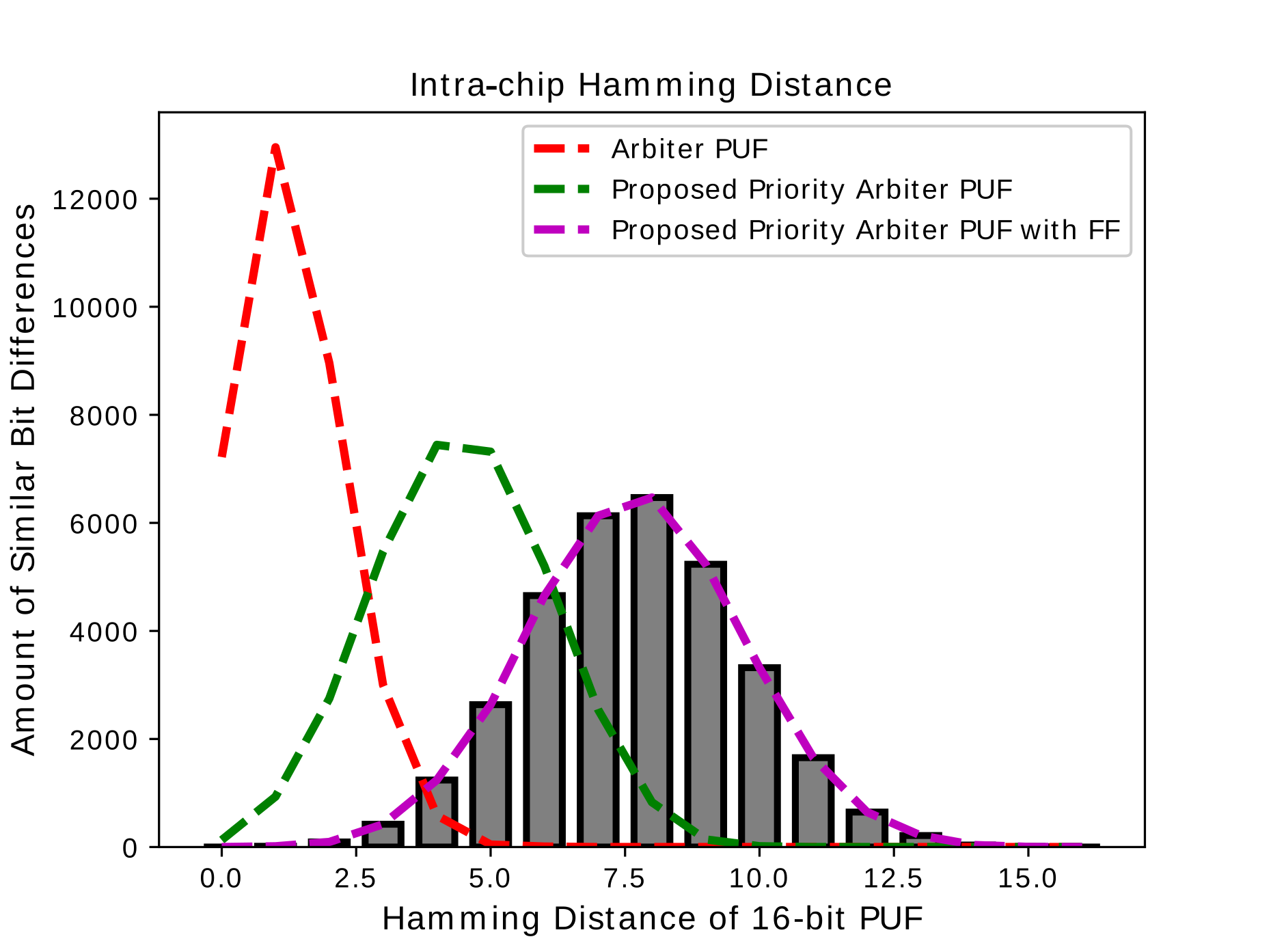}
\caption{Hamming distance of the proposed PA-PUF with and without feed-forward arbiters and the classical arbiter PUF. \vspace{-3mm}}
\label{fig:HD16_comp}
\end{figure}

\subsection{Machine Learning-based Modelling Attacks}
The key purpose of PUFs can be defeated by modeling the PUF structure and by being able to predict its output. In a series of works~\cite{pufml1, pufml2, pufml3}, it has been shown that nearly all variants of delay-based PUFs can be efficiently modeled using machine learning. To further boost security, XOR-based arbiter PUFs are introduced. It was also shown in a recent study~\cite{pufml4} that learning XOR-based arbiter PUFs is possible up to a limit on the number of parallel arbiter chains. Complementing this research direction, alternative PUF design frameworks~\cite{pufml5} or restricted visibility of the PUF outputs~\cite{pufml6} are proposed. Since our proposed design applies to the general studies on arbiter-based PUFs, it will come under the same purview of attacks and resilience presented earlier. Hence, we focus on the lightweight PUF design itself and reserve the study of detailed analysis of modeling attacks for the future. In that context, it will be interesting to juxtapose the priority arbiter design against the XOR-based arbiter chain merger.

\section{Comparisons}
\label{sec:comp}
Based on the performance metrics calculation, the comparisons of the proposed PUF with the existing designs is summarized in Table \ref{tab:comp}. It is evident from the results that the proposed priority arbiter PUF has good results compared to the existing designs.

\begin{figure}[!t]
\centering
\includegraphics[scale=0.52]{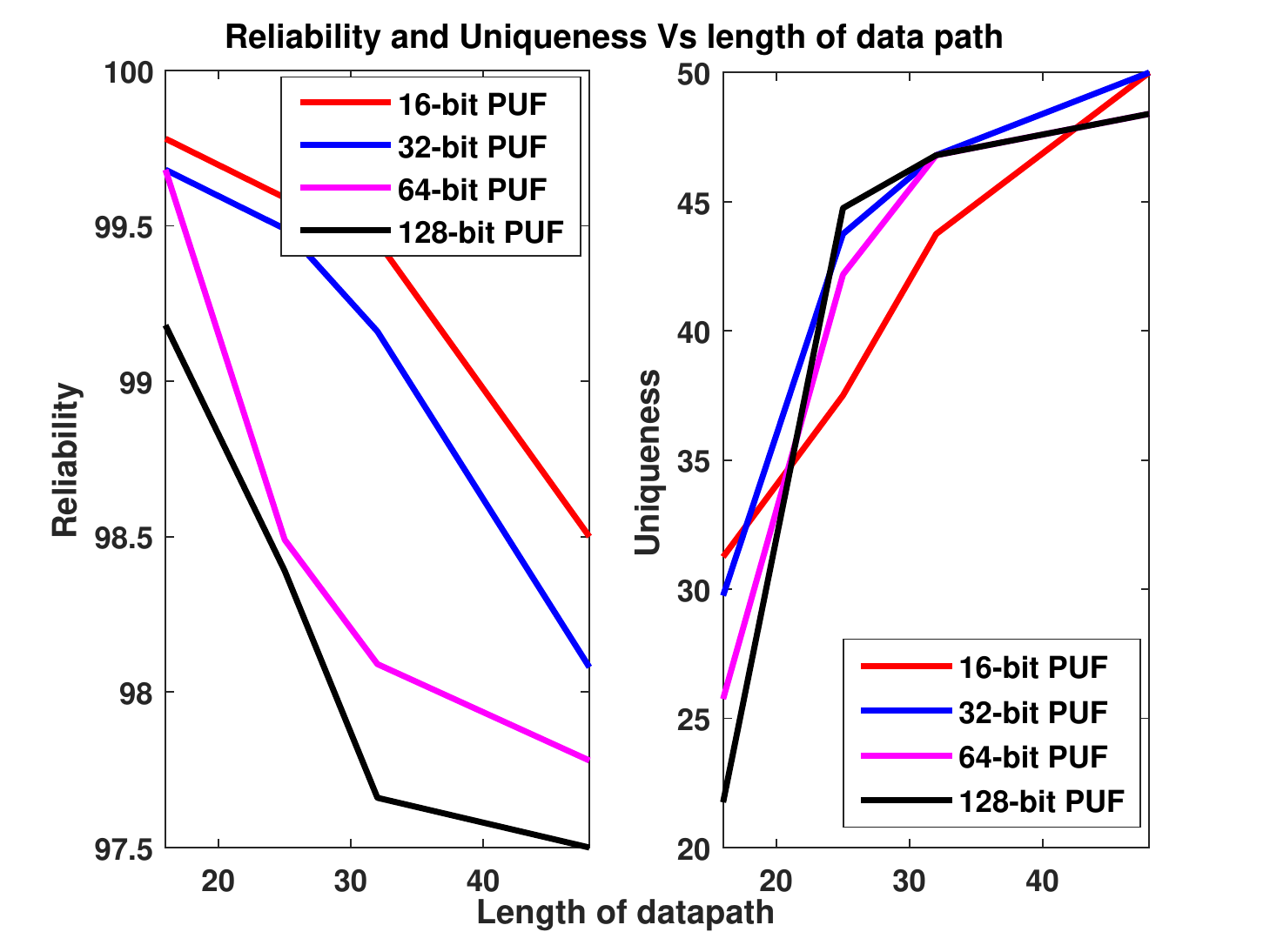}
\caption{Comparison of {\em uniqueness} and {\em reliability} with respect to data path's length (16-bit challenge + feed-forward~arbiters).\vspace{-3mm}}
\label{fig:uniq:rel_cmp_datapath}
\end{figure}
  
\begin{table}[!b]
\centering
\setlength{\extrarowheight}{3.0pt}
\scriptsize
\caption{Comparison of the performance metrics of the ring oscillator PUF; Ideal value of reliability is 100$\%$  while the  remaining parameters have an ideal value of 50$\%$.}
    \begin{tabular}{|c|c|c|c|c|c|}
    \hline
    \textbf{Parameter}        & \textbf{Proposed}  & \textbf{RO}         	& \textbf{Original}       & \textbf{CHAR}   & \textbf{CHAR\& MAJ}               \\  \cline{4-6}
    \textbf{\%}  		    &  \textbf{PA-PUF}  &\textbf{PUF}  	 &  \multicolumn{3}{c|}{\textbf{Design}  \cite{GuNeill2017a}} \\     \hline
    Uniqueness      &  49.63 		& 49.22       &     48.52      & 45.60 			& 45.60          \\ \hline
    Uniformity       & 	49.45			& 48.5          & 51.06 		   & 50.60 			& 50.54              \\ \hline
    Reliability        & 100				& 99.99      & 92.00        & 98.87   			& 99.58          \\ \hline
    Bit-aliasing    	& 	49.66			&47.89        & 43.52       &  43.52 			& 43.52     \\ \hline
    \end{tabular}
     \label{tab:comp}
\end{table}

The length of the data path plays a prominent role in deriving the response. In addition to the
length of the multiplexer-data path, the number of feed-forward arbiters also plays a key role. Fig.~\ref{fig:HD16_comp}  shows the comparison of the classical arbiter PUF, proposed priority arbiter PUF, and the priority arbiter PUF with the feed-forward arbiter. The plot reveals that the classical arbiter PUF is not able to 
produce a various number of responses while the proposed designs have more number of CRPs. Since the intra-Hamming distance plot has a maximum at half of its response length only for the priority arbiter with the feed-forward arbiter. The number of CRPs in a PA-PUF can be increased by using the feed-forward arbiters.

Further, the number of feed-forward arbiters used as a select signal to the multiplexer in the data path also influences the 
response. The plot shown in Fig.~\ref{fig:uniq:rel_cmp_datapath} reveals that as the number of feed-forward arbiters increases in the design,
the reliability of the PUF reduces and the uniqueness of the PUF improves. 
Since the number of CRPs can be increased by increasing the data path length and by introducing more feed-forward arbiters, the PUF is able to generate more variations in the output. Hence, the reliability is going to reduce.
So, one can decide the number of feed-forward arbiters as per their requirements of uniqueness and reliability. 

\begin{table}[!b]
\scriptsize
\centering
\caption{Comparisons of different PUF designs; `U' is the uniqueness and `R' is the reliability.}
\begin{tabular}{|c|c|c|c|c|} \hline


\makecell{\textbf{PUF design} $^{(\textbf{response~size})}$}			& \textbf{U} (\textbf{\%}) 				 	& \textbf{R} \textbf{(\%)} 			&  \textbf{Platform}	& \textbf{Area}   \\  
\hline
\makecell{\textbf{Proposed PA-PUF} $^{\textbf{(128)}}$}  & $>$ \textbf{49.63} 			   & \textbf{100}          & \textbf{Artix-7}                     			  &\makecell{\textbf{47 slices} \\ \textbf{per PUF}}  \\
\hline




\makecell{SRAM PUF \cite{Guajardo2007a} $^{(128)}$} & 49.97 & $>$ 88 & FPGA & \makecell{4800 SRAM \\ memory bits} \\
\hline

\makecell{Latch PUF \cite{Yamamoto2011a} \\ $^{(128)}$} 		& 46    		&  $>$  87   & Spartan-3               			& 2$\times$128 slices \\
\hline

\makecell{Flip flop PUF \cite{Maes2008a} \\ $^{(4096)}$}  		& $\approx$50    	&  $>$ 95    & Virtex-2              		& \makecell{4096 \\flip-flops} \\
\hline

\makecell{Butterfly PUF \cite{kumar2008a} \\ $^{(64)}$ } 		& $\approx$50      & 94    & Virtex-5                 			& 130 slices \\
\hline
 
\makecell{RO PUF \cite{SuhDevadas2007a}\\  $^{(128)}$} 			& 46.15     &     99.52     & Virtex-4               					& 1024 RO's \\
\hline

\makecell{CRO PUF \cite{Merli2010a} \\ $^{(127)}$}  			& 43.50    & 96   		& Spartan-3                					& 64 slices \\
\hline

\makecell{PUF ID  \cite{Gu2014a} \\ $^{(128)}$ }   							& 49.90     &93.93    & Artix-7              				& 128 slices \\
\hline


\makecell{Ultra-compact PUF ID \cite{Gu2015a}\\ $^{(128)}$}    									& 49.93    &  93.96      & Spartan-6              				& 40 slices \\
 \hline

\makecell{Improved PUF ID \cite{GuNeill2017a} \\$^{(128)}$  }			  & 45.60     		  & 99.42   &Spartan-6              		    	&128 slices \\
 \hline

\end{tabular}
\label{tab:comp_diff_puf}
\end{table}

\section{Conclusions}
\label{sec:conclusions}

This paper proposed a new priority arbiter with a uniform output with an equal probability of 0's and 1's. We also proposed a new priority arbiter-based PUF (PA-PUF) with three data path lines and a priority arbiter. Further, the feed-forward arbiters are introduced to get new select signals to the multiplexers in the data path to improve the number of challenge-response pairs. Finally, we demonstrated the results of the proposed PA-PUF on Nexys Video Artix-7 FPGA board.
The experimental results show that the proposed PA-PUF has a reliability of 94.5$\%$, which can be improved to 100\% by implementing error correction techniques. Moreover, the PA-PUF improves the PUF uniformity to 49.45\% and uniqueness to 49.63$\%$. We plan to study the attacks on PA-PUF in more detail, practically and theoretically, in the future.


\small{
\bibliographystyle{IEEEtran}
\bibliography{ref}
}

\end{document}